\documentclass[prd,draft,preprint,showpacs,groupedaddress]{revtex4-1}
\usepackage{amsmath}
\begin{document}


\title{Generalized holographic equipartition for Friedmann-Robertson-Walker
universes}


\author{Wen-Yuan Ai}
\author{Hua Chen}
\author{Xian-Ru Hu}
\author{Jian-Bo Deng}

\email[Jian-Bo Deng: ]{dengjb@lzu.edu.cn}

\affiliation{Institute of Theoretical Physics, LanZhou University, Lanzhou 730000, P. R. China}


\date{\today}

\begin{abstract}
The novel idea that spatial expansion of our universe can be regarded as the consequence of the emergence of space was proposed by Padmanabhan. By using of the basic law governing the emergence, which Padmanabhan called holographic equipartition, he also arrives at the Friedmann equation in a flat universe. When generalized to other gravity theories, the holographic equipartition need to be generalized with an expression of $f(\Delta N,N_{sur})$. In this paper, we give general expressions of $f(\Delta N,N_{sur})$ for generalized holographic equipartition which can be used to derive the Friedmann equations of the Friedmann-Robertson-Walker universe with any spatial curvature in higher (n+1)-dimensional Einstein gravity, Gauss-Bonnet gravity and more general Lovelock gravity. The results support the viability of the perspective of holographic equipartition.
\end{abstract}



\maketitle

\section{INTRODUCTION}
Since the discovery of black hole thermodynamics \cite{bardeen1973four, hawking1975particle} in the 1970s, we have got a lot of insights of the nature of gravity from black hole physics. The tight connections between gravitation and thermodynamics have been fully studied. By considering the entropy bound of gravitational system, a fundamental principle has been built, called holographic principle \cite{bousso2002holographic}. The holographic principle states that the degrees of freedom (DOF) of a gravitational system are determined by the area of the boundary not the volume of the bulk \cite{bousso2002holographic}. And it was first implemented by the dual ADS/CFT \cite{maldacena1999large, witten1998anti}. It implies that gravity may not be a fundamental interaction but an emergent phenomenon.\par
Jacobson \cite{jacobson1995thermodynamics} was the first one who showed that how Einstein field equations can be derived from the Clausius relation on a local Rindler causal horizon. Verlinde \cite{verlinde2011origin} took a great step forward by suggesting that the gravity may not be a fundamental interaction but should be explained as an entropic force caused by changes of entropy associated with the information on the holographic screen. With the holographic principle and the equipartition law of energy, the Newton's law of gravitation was derived by Verlinde, moreover, in a relativistic regime the Einstein equations were also derived. Earlier, Padmanabhan \cite{padmanabhan2010equipartition} also observed that equipartition law of energy for the horizon DOF combining with the thermodynamic relation $S=E/2T$, leads to the Newton's law of gravity.  See Refs. \cite{padmanabhan2010thermodynamical, padmanabhan2011lessons} for a review.\par
However in the most cases, one only treats the gravitational field as an emergent phenomenon, but assuming the existence of spacetime manifold. It is obviously a more complete way to view the emergence of gravitational phenomenon that the spacetime is also regarded as an emergent structure. The question if the spacetime itself is an emergent structure was answered by Padmanabhan \cite{padmanabhan2012emergence} recently. Though there are some conceptual difficulties associated with this idea, the answer is positive when one considers the emergence of spacetime in cosmology. Padmanabhan argued that our universe provides a setup to stress the issue that \emph{the cosmic space is emergent as the cosmic time progresses}. He argued that the spatial expansion of our universe is due to the difference between the surface DOF and the bulk DOF in a region of emerged space and finally derived the Friedmann equation of a flat Friedmann-Robertson-Walker (FRW) universe.\par
Cai \cite{cai2012emergence} generalized the derivation process to the higher (n+1)-dimensional spacetime. He also obtained the Friedmann equations of a flat FRW universe in Gauss-Bonnet and more general Lovelock cosmology by properly modifying the effective volume and the number of DOF on the holographic surface from the entropy formulas of static spherically symmetric black holes \cite{cai2012emergence}. In Ref. \cite{yang2012emergence}, on the other hand, the authors generalized the holographic equipartition and derived the Friedmann equations by assuming that $(dV/dt)$ is proportional to a general function $f(\Delta N,N_{sur})$. Here $\Delta N=N_{sur}-N_{bulk}$, where $N_{sur}$ and $N_{bulk}$ are, respectively, the number of DOF on the boundary and in the bulk. However, though the authors of \cite{cai2012emergence, yang2012emergence} can obtain the Friemann equation of the FRW universe with spacial curvature in Einstein gravity, they failed to derive Friedmann equations of a nonflat FRW universe corresponding to the Gauss-Bonnet gravity and the Lovelock gravity. For more investigations about the novel idea see Refs. \cite{sheykhi2013friedmann, tu2013emergence, ai2013emergence}.\par
In this paper, we will give general expressions of $f(\Delta N,N_{sur})$ for generalized holographic equipartition which can be used to derive the Friedmann equations in the FRW universe with any spatial curvature in higher (n+1)-dimensional Einstein gravity, Gauss-Bonnet gravity and Lovelock gravity. Next section, a brief review of Padmanabhan and others' work will be firstly presented. In \mbox{Section \uppercase\expandafter{\romannumeral3}} we will give general expressions of $f(\Delta N,N_{sur})$ and show that how Friedmann equations in the FRW universe with any spatial curvature can be derived from these expressions. \mbox{Section \uppercase\expandafter{\romannumeral4}} is for discussions and conclusions.\par

\section{EMERGENCE OF SPACE AND THE HOLOGRAPHIC EQUIPARTITION}
First, let us recall Padmanabhan's observation \cite{padmanabhan2012emergence}. He notices that in a pure de Sitter universe with Hubble constant H, the holographic principle can be expressed in terms of
\begin{equation}
\label{1} N_{sur}=N_{bulk},
\end{equation}
where $N_{sur}$ denotes the number of DOF on the spherical surface of Hubble radius $ H^{-1} $, $ N_{sur}=4\pi H^{-2}/L_p^2 $, with $ L_p $ being the Planck length, while the bulk DOF $ N_{bulk}=\left|E\right|/(1/2)T $. Here $ \left|E\right|=\left|\rho+3p\right|V $, is the Komar energy and the horizon temperature $ T=H/2\pi $. For the pure de Sitter universe, substituting $ \rho=-p $ into Eq. \eqref{1}, the standard result $ H^2=8\pi L_p^2\rho/3 $ is obtained.\par
From Eq. \eqref{1}, one has $|E|=(1/2)N_{sur}T$, which is the standard equipartition law. Since it relates the effective DOF residing in the bulk to the DOF on the boundary surface, Padmanabhan called it \emph{holographic equipartition}. It is known that our real universe is just asymptotically de Sitter. Padmanabhan further suggested that the emergence of space occurs and relates to the difference $\Delta N=N_{sur}-N_{bulk}$. A simple equation was proposed \cite{padmanabhan2012emergence}
\begin{equation}
\label{2} \frac{dV}{dt}=L_p^2\Delta N.
\end{equation}
Putting the above definition of each term, one obtains
\begin{equation}
\label{3} \frac{\ddot{a}}{a}=-\frac{4\pi L_p^2}{3}(\rho +3p).
\end{equation}
This is the standard dynamical equation for the FRW universe in general relativity. Using continuity equation $\overset{.}{\rho}+3H(\rho+p)=0$, one gets the standard Friedmann equation
\begin{equation}
\label{4} H^2+\frac{k}{a^2}=\frac{8\pi L_p^2\rho}{3},
\end{equation}
where k is an integration constant, which can be interpreted as the spatial curvature of the FRW universe. Here, Padmanabhan takes $(\rho+3p)<0$, which makes sense only in the accelerating phase. It means that in order to have the asymptotic holographic equipartition, the existence of dark energy is necessary.\par
Based on Cai's work \cite{cai2012emergence}, which generalizes the derivation process to the higher (n+1)-dimensional spacetime, the authors of \cite{yang2012emergence} generalized the holographic equipartition. They assumed that $(dV/dt)$ is proportional to a general function $(\Delta N,N_{sur})$, i.e.
\begin{equation}
\label{5} \frac{dV}{dt}=L_p^{n-1}f(\Delta N,N_{sur}).
\end{equation}
Instead of modifying the number of DOF on the surface of the Hubble sphere and the volume increase, they also derived Friedmann equations of the spatially flat FRW universe in Gauss-Bonnet and Lovelock gravities from the generalized law governing the emergence of cosmic space \eqref{5} with different expressions of $f(\Delta N,N_{sur})$ relate to different gravity theories. \par

\section{GENERAL EXPRESSIONS OF $f(\Delta N,N_{sur})$ FOR GENERALIZED HOLOGRAPHIC EQUIPARTITION}
In this section, we will give general expressions of $f(\Delta N,N_{sur})$ which can be used to derive the Friedmann equations of the FRW universe with any spatial curvature in higher (n+1)-dimensional Einstein gravity, Gauss-Bonnet gravity and Lovelock gravity. As in \cite{sheykhi2013friedmann}, we will apply Eq. \eqref{5} to the apparent horizon instead of the Hubble sphere. Under the cosmological principle, we have the (n+1)-dimensional FRW metric
\begin{equation}
\label{6} ds^2=-dt^2+a(t)^2\left(\frac{dr^2}{1-kr^2}+r^2d\Omega_{n-1}^2\right),
\end{equation}
where k denotes the spatial curvature constant with k=+1, 0 and -1 corresponding to a closed, flat and open universe, respectively. $d\Omega _{n-1}^2$ is the line element of a (n-1)-dimensional unite sphere. Adopting a new coordinate $\tilde{r}=a(t)r$, the metric \eqref{6} will become
\begin{equation}
\label{7} ds^2=h_{ab}dx^adx^b+\tilde{r}d\Omega_{n-1}^2,
\end{equation}
with $x^0=t$, $x^1=r$, $h_{ab}=diag(-1,a^2/(1-kr^2 ))$. The dynamical apparent horizon $\tilde{r}_A$ is determined by $h^{ab}\partial_a\tilde{r}\partial_b\tilde{r}=0$, and is given by
\begin{equation}
\label{8} \tilde{r}_A=\frac{1}{\sqrt{H^2+\frac{k}{a^2}}}.
\end{equation}
The number of DOF on the apparent horizon is \cite{verlinde2011origin}
\begin{equation}
\label{9} N_{sur}=\frac{\alpha A}{L_p^{n-1}},
\end{equation}
where
\begin{equation}
\label{10} A=n\Omega_n\tilde{r}_A^{n-1},
\end{equation}
and $\alpha=(n-1)/2(n-2)$ with $\Omega_n$ being the volume of an n sphere of unite radius. The number of DOF inside the apparent horizon is given as
\begin{equation}
\label{11} N_{bulk}=\frac{|E|}{T/2},
\end{equation}
where $E=[(n-2)\rho+np]V/(n-2)$ is the bulk Komar energy \cite{cai2010friedmann}. $V=\Omega_n\tilde{r}_A^n$ is the bulk volume and the temperature of the apparent is given by $T=1/2\pi\tilde{r}_A$. As in Refs. \cite{cai2012emergence, yang2012emergence}, here only the accelerating phase with $(n-1)\rho+np<0$ is considered.\par
In the (n+1)-dimensional Einstein gravity, we choose the $f(\Delta N,N_{sur})$ as
\begin{equation}
\label{12} f(\Delta N,N_{sur})=\left[1-\frac{k}{a^2\lambda}\left(\frac{N_{sur}}{\alpha}\right)^{\frac{2}{n-1}}\right]^{\frac{1}{2}}\frac{\Delta N}{\alpha},
\end{equation}
where $\lambda=(n\Omega_n/L_p^{n-1})^{2/(n-1)}$, k is the spatial curvature constant. Substituting the above expressions of V, $N_{sur}$ and $N_{bulk}$ into Eq. \eqref{5}, one gets
\begin{equation}
\label{13} \tilde{r}_A^{-2}-\dot{\tilde{r}}H^{-1}\tilde{r}_A^{-3}=-\frac{8\pi L_p^{n-1}}{n(n-1)}[(n-2)\rho +np].
\end{equation}
Multiplying the both hand side by factor $2\dot{a}a$, after using the continuity equation in (n+1)-dimensional
\begin{equation}
\label{14} \dot{\rho}+nH(\rho +p)=0
\end{equation}
and relation \eqref{8}, we obtain
\begin{equation}
\label{15} \frac{dV}{dt}\left[a^2\left(H^2+\frac{k}{a^2}\right)\right]=\frac{16\pi L_p^{n-1}}{n(n-1)}\frac{d}{dt}(\rho a^2).
\end{equation}
Integrating, the final Friedmann equation of the (n+1)-dimensional FRW universe with any spatial curvature
\begin{equation}
\label{16} H^2+\frac{k}{a^2}=\frac{16\pi L_p^{n-1}}{n(n-1)}\rho,
\end{equation}
is derived, where the integration constant is set to be zero.\par

Then,the $f(\Delta N,N_{sur})$ in Gauss-Bonnet gravity is chosen as
\begin{equation}
\label{17} f(\Delta N,N_{sur})=\left[1-\frac{k}{a^2\lambda}\left(\frac{N_{sur}}{\alpha}\right)^{\frac{2}{n-1}}\right]^{\frac{1}{2}}\frac{\frac{\Delta N}{\alpha}+\tilde{\alpha}\lambda(N_{sur}/\alpha)^{1+\frac{2}{1-n}}}{1+2\tilde{\alpha}\lambda (N_{sur}/\alpha)^{\frac{2}{1-n}}},
\end{equation}
where $\tilde{\alpha}$ is the Gauss-Bonnet coefficient \cite{cai2005first}. When $\tilde{\alpha}=0$, this choice \eqref{17} reduces to the previous one \eqref{12}. Substituting Eq. \eqref{17} into Eq. \eqref{5}, we have
\begin{equation}
\label{18} (\tilde{r}_A^{-2}+\tilde{\alpha}\tilde{r}_A^{-4})-\dot{\tilde{r}}_AH^{-1}\tilde{r}_A^{-3}(1+2\tilde{\alpha}\tilde{r}_A^{-2})=-\frac{8\pi L_p^{n-1}}{n(n-1)}[(n-2)\rho +np].
\end{equation}
Multiplying the both hand side of \eqref{18} by factor $2\dot{a}a$, with the help of continuity equation \eqref{14} and relation \eqref{8}, one gets
\begin{equation}
\label{19} \frac{d}{dt}\left\{a^2\left[H^2+\frac{k}{a^2}+\tilde{\alpha}\left(H^2+\frac{k}{a^2}\right)^2\right]\right\}=\frac{16\pi L_p^{n-1}}{n(n-1)}\frac{d}{dt}(\rho a^2).
\end{equation}
Integrating, we obtain
\begin{equation}
\label{20}H^2+\frac{k}{a^2}+\tilde{\alpha}\left(H^2+\frac{k}{a^2}\right)^2=\frac{16\pi L_p^{n-1}}{n(n-1)}\rho,
\end{equation}
where again the integration constant is set to be zero. This is nothing but the corresponding Friedmann equation of the FRW universe with any spatial curvature in Gauss-Bonnet gravity \cite{cai2005first}.

In the case of Lovelock gravity, a more general expression of $f(\Delta N,N_{sur})$ is given as
\begin{equation}
\label{21}f(\Delta N,N_{sur})=\left[1-\frac{k}{a^2\lambda}\left(\frac{N_{sur}}{\alpha}\right)^{\frac{2}{n-1}}\right]^{\frac{1}{2}}\frac{\frac{\Delta N}{\alpha}+\Sigma^m_{i=2}\hat{c}_i\lambda_i(N_{sur}/\alpha)^{1+\frac{2i-2}{1-n}}}{1+\Sigma^m_{i=2}i\hat{c}_i\lambda_i(N_{sur}/\alpha)^{\frac{2i-2}{1-n}}},
\end{equation}
where $\lambda_i=(n\Omega_n/L_p^{n-1})^{(2i-2)/(n-1)}$, $m=[n/2]$. $\hat{c}_i$ are some coefficients in front of Euler density terms with dimension (2i-1), and especially $\hat{c}_1=1$. If $\hat{c}_i=0$ for $i>2$, then function \eqref{21} recovers the Gauss-Bonnet one. Substituting Eq. \eqref{21} into Eq. \eqref{5}, we have
\begin{equation}
\label{22}\overset{m}{\underset{i=1}{\Sigma}}\hat{c}_i\tilde{r}_A^{-2i}-\dot{\tilde{r}}_AH^{-1}\overset{m}{\underset{i=1}{\Sigma}}i\hat{c}_i\tilde{r}_A^{-2i-1}=-\frac{8\pi L_p^{n-1}}{n(n-1)}[(n-2)\rho +np].
\end{equation}
As previously, multiplying both hand side of \eqref{22} by factor $2\dot{a}a$, after using the continuity equation \eqref{14} and relation \eqref{8}, we get
\begin{equation}
\label{23}\frac{d}{dt}\left[a^2\overset{m}{\underset{i=1}{\Sigma}}\hat{c}_i\left(H^2+\frac{k}{a^2}\right)^i\right]=\frac{16\pi L_p^{n-1}}{n(n-1)}\frac{d}{dt}(\rho a^2).
\end{equation}
After integrating and setting the integration constant to zero ,we finally obtain
\begin{equation}
\label{24}\overset{m}{\underset{i=1}{\Sigma}}\hat{c}_i\left(H^2+\frac{k}{a^2}\right)^i=\frac{16\pi L_p^{n-1}}{n(n-1)}\rho.
\end{equation}
This is indeed the Friedmann equation of the FRW universe in Lovelock theory \cite{cai2005first}.

Up to now, we have got all the general expressions of $f(\Delta N,N_{sur})$, which can be used to derive the Friedmann equations in higher (n+1)-dimensional Einstein gravity, Gauss-Bonnet gravity and more general Lovelock gravity. We note that in order to get the Friedmann equations of the FRW universe with any curvature, we actually need to apply Eq. \eqref{5} to the apparent horizon rather than the Hubble horizon. As a consequence, even in (3+1)-dimensional Einstein gravity, the expression of $f(\Delta N,N_{sur})$ need to be modified by $ f(\Delta N,N_{sur})=\Delta N\sqrt{1-(kN_{sur}/a^2\lambda)} $, which comes from Eq. \eqref{12}. It is clearly very different from the one in Ref. \cite{yang2012emergence}.

\section{CONCLUSION AND DISCUSSION}
To summarize, in this paper, the novel idea made by Padmanabhan \cite{padmanabhan2012emergence} that the emergence of space and expansion of the universe are due to the difference between the number of DOF on the holographic surface and the one in the emerged bulk is investigated. It is shown that the Friedmann equation of a flat FRW universe can be derived  with the help of continuity equation \cite{padmanabhan2012emergence}. However when one generalizes the derivation process to general gravity theories, one need to modify the number of DOF on the surface of the Hubble sphere and the volume increase \cite{cai2012emergence}, or another way, with $\Delta N$ taken place by a general function of $(\Delta N,N_{sur})$ \cite{yang2012emergence}. But the authors of Ref. \cite{yang2012emergence} derived the Friedmann equations of only a flat FRW universe in Gauss-Bonnet and Lovelock gravities by expressions of $f(\Delta N,N_{sur})$ given in Ref. \cite{yang2012emergence}. In this paper we gave general expressions $f(\Delta N,N_{sur})$ of generalized holographic equipartition which can be used to derive Friedmann equations of the FRW universe with any spatial curvature in higher dimensional general relativity, Gauss-Bonnet gravity and Lovelock gravity. The key point in doing this is applying Eq. \eqref{5} to the apparent horizon. Though we can obtain the Friedmann equation of the FRW universe with spacial curvature in Einstein gravity when Eq. \eqref{5} is applied to the Hubble horizon. Only when it is applied to the apparent horizon, can we arrive at the Friedmann equations of the FRW universe with any spacial curvature in Gauss-Bonnet and Lovelock gravities. Thus it is better to think that the holographic equipartition actually holds for the apparent horizon rather than the Hubble horizon.

\bibliography{refrence}

\end{document}